\documentclass[sn-mathphys,Numbered]{sn-jnl}

\usepackage{graphicx}%
\usepackage{multirow}%
\usepackage{amsmath,amssymb,amsfonts}%
\usepackage{amsthm}%
\usepackage{mathrsfs}%
\usepackage[title]{appendix}%
\usepackage{xcolor}%
\usepackage{textcomp}%
\usepackage{manyfoot}%
\usepackage{booktabs}%
\usepackage{algorithm}%
\usepackage{algorithmicx}%
\usepackage{algpseudocode}%
\usepackage{listings}%

\theoremstyle{thmstyleone}%
\newtheorem{theorem}{Theorem}
\newtheorem{proposition}[theorem]{Proposition}%

\theoremstyle{thmstyletwo}%
\newtheorem{example}{Example}%
\newtheorem{remark}{Remark}%

\theoremstyle{thmstylethree}%
\newtheorem{definition}{Definition}%

\begin{document}

\title[Forecasting Soccer Matches through Distributions]{Forecasting Soccer Matches through Distributions}


\author*[1,2]{\fnm{Tiago} \sur{Mendes-Neves}}\email{tiago.m.neves@inesctec.pt}

\author[1,2]{\fnm{Yassine} \sur{Baghoussi}}\email{baghoussi@fe.up.pt}

\author[3]{\fnm{Luís} \sur{Meireles}}\email{luiscunhameireles1@gmail.com}

\author[1,4]{\fnm{Carlos} \sur{Soares}}\email{csoares@fe.up.pt}

\author[1,2]{\fnm{João} \sur{Mendes-Moreira}}\email{jmoreira@up.pt}

\affil*[1]{\orgdiv{Faculdade de Engenharia}, \orgname{Universidade do Porto}, \orgaddress{\city{Porto}, \country{Portugal}}}

\affil[2]{\orgdiv{LIAAD}, \orgname{INESC TEC}, \orgaddress{\city{Porto}, \country{Portugal}}}

\affil[3]{\orgname{Nordensa Football}, \orgaddress{\city{Cluj}, \country{Romania}}}

\affil[4]{\orgname{Fraunhofer AICOS Portugal}, \orgaddress{\city{Porto}, \country{Portugal}}}


\abstract{

Forecasting sporting events encapsulate a compelling intellectual endeavor, underscored by the substantial financial activity of an estimated \$80 billion wagered in global sports betting during 2022, a trend that grows yearly.
Motivated by the challenges set forth in the Springer Soccer Prediction Challenge, this study presents a method for forecasting soccer match outcomes by forecasting the shot quantity and quality distributions. The methodology integrates established ELO ratings with machine learning models.
The empirical findings reveal that, despite the constraints of the challenge, this approach yields positive returns, taking advantage of the established market odds.

}




\maketitle

\section{Introduction}

The sport of soccer, also known as (association) football, is underpinned by a complex network of factors that determine the trajectory of each match. This inherent complexity presents a fascinating and challenging task when forecasting match outcomes. The Springer Soccer Prediction Challenge provides a platform to tackle this problem.

Forecasting soccer games is challenging due to several factors, including the high prevalence of draws \cite{bunker_application_2022}, the unpredictability of individual matches, and the complex interactions between teams, players, and coaches. Additionally, while detailed data exists in some scenarios (e.g., event and tracking), in some cases, like the Springer Soccer Prediction Challenge, only basic statistics can be used (e.g., results). While results data is easier to collect, it lacks the intricate detail that might be available in more granular sources. Nevertheless, this constraint prompts us to explore the essence of the game from a different perspective and compels us to approach the problem with innovative tactics. In an era characterized by an avalanche of data, such challenges provide a critical opportunity for reflection and advancement in machine learning fundamentals.

We offer an alternative approach to soccer results forecasting. Instead of forecasting the probability of a win or the number of goals each team is likely to score, we focus on forecasting the shot quantity and quality distribution for each team. The goal is to sample a list of expected goals, which, through simulation, can be used to forecast the probabilities and correct scores, similar to concepts such as Expected Points.

\begin{figure}[h]
\centering
\includegraphics[width=0.75\textwidth]{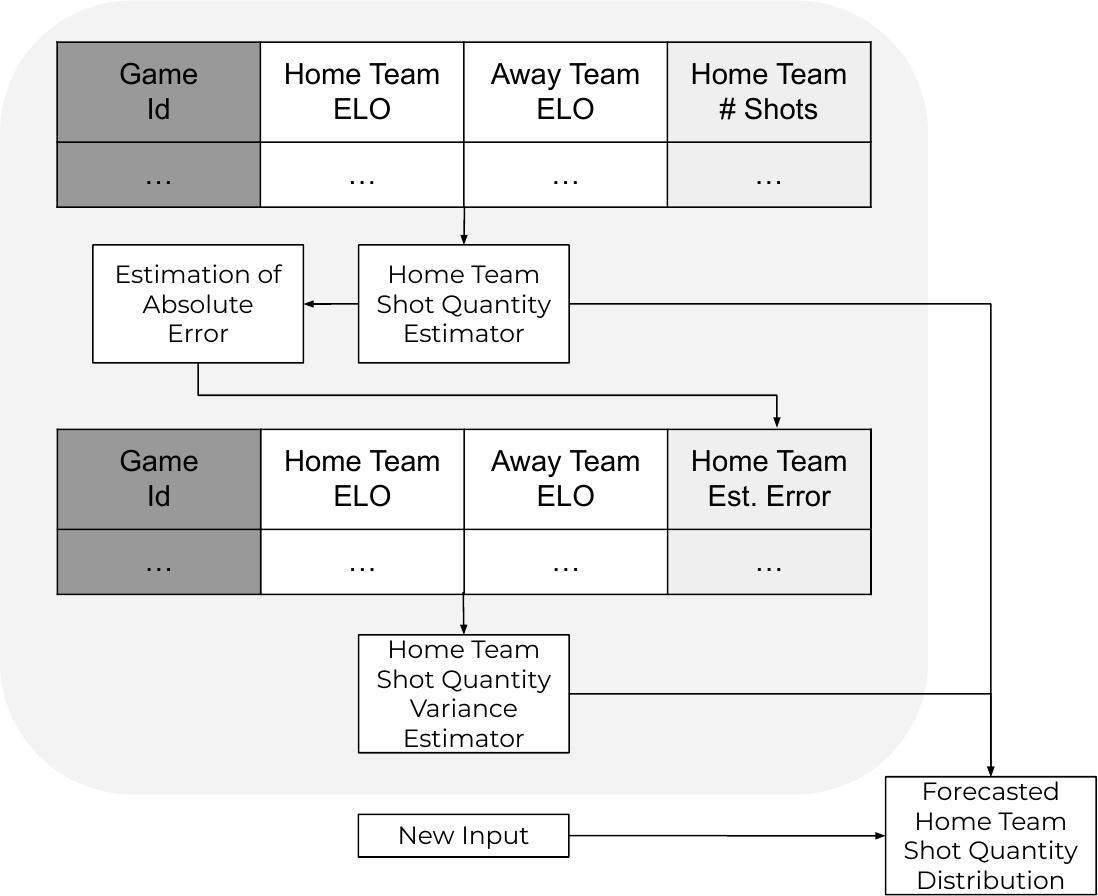}
\caption{Our method to forecast the distributions of relevant variables (in this example, the number of shots made for the home team). First, we build an estimator for the average value of the distribution. Then, using the instance-wise error of the estimator, we build another estimator that forecasts the expected variance of the distribution. By assuming a normal distribution, we can estimate the distribution for any new input. Steps in the grey area use training data only. The model ignores dark-grey columns, while light-grey columns represent the target variable.}\label{fig:methods_intro}
\end{figure}

Our methodology unfolds in several steps. First, we calculate the ELO ratings for each team, which serve as indicators of their relative strengths. Next, we train two linear models: one for the number of shots in a game and another for the expected goals per shot. These linear models use both teams' ELO ratings to make forecasts. The prediction errors in the training set are then used to model the expected variance. We illustrate the procedure in Figure \ref{fig:methods_intro}.

Assuming a normal distribution, we forecast each team's shot quantity and quality distribution. Using these distributions, we sample the number of shots and the quality of each shot, leading to a list of expected goals for each team. Multiple instances of the game are simulated using these expected goals. Finally, we predict probabilities based on the frequency of an event relative to the total number of simulations.

The rest of this paper is organized as follows. Section \ref{sec:literature_review} discusses the related work on soccer forecasting. Section \ref{sec:formulation} formulates our solution. Section \ref{sec:method} describes our method in detail. Section \ref{sec:experiments} provides information on the experiments, namely for the selection of the method to estimate distributions. Section \ref{sec:results} presents our results. Section \ref{sec:discussion} presents the discussion of the limitations of our study and future research directions.

\section{Literature Review} \label{sec:literature_review}

Soccer match forecasting methodologies have seen significant advancements, using traditional statistical techniques and modern machine learning algorithms. One well-established methodology is based on the Poisson distribution model \cite{karlis_analysis_2003}, a probability model that predicts the number of goals scored by each team. This model assumes that the number of goals scored by a team in a soccer match follows a Poisson distribution, and factors such as a team's offensive and defensive strengths, the team's form, and home ground advantage, among others, are integrated into the model. While the Poisson model has seen widespread usage, its assumptions may not always hold, particularly given the complex and unpredictable nature of soccer matches.

An alternative approach is the ELO rating system \cite{arpad_e_elo_proposed_1967}, a methodology originally designed for chess but has found application in soccer prediction \cite{hvattum_using_2010}. ELO ratings reflect a team's strength based on their past performance. The system awards points for match wins, with more points awarded for winning against stronger opponents and vice versa. Over time, a team's ELO rating will adjust to reflect their true skill level. The ELO model is often used to predict the outcome of a match by comparing the ELO ratings of the two teams. While the ELO system provides a robust and dynamic means of forecasting, it does have its limitations, such as not considering recent team form or the impact of individual player performance. Furthermore, it also doesn't account for factors such as injuries or suspensions, which can significantly influence a match outcome. Nonetheless, ELO inspired many rating systems such as Pi-rating \cite{constantinou_determining_2013} and other similar approaches \cite{wheatcroft_profitable_2020,constantinou_dolores_2019}.
The paper \cite{lazova_pagerank_2015} applies the PageRank with restarts algorithm to a graph built from the games played during the tournaments to rank national teams based on all matches during the championships.

The literature significantly relies on statistical methods in predicting football match outcomes. Regression models for forecasting goals and match results is a common approach \cite{goddard_regression_2005,dixon_modelling_1997}.

Machine learning models have been increasingly used in sports betting predictions \cite{dubitzky_open_2019}. \cite{hubacek_learning_2019} describes a winning solution to the 2017 Springer Soccer Prediction Challenge using gradient-boosted trees.
\cite{helic_comparing_2020} compares neural network ensemble approaches in soccer predictions, showing that ensemble methods increase accuracy and rentability over the best single model.
In machine learning models, the literature emphasizes the importance of generating insightful features and domain knowledge in improving predictions \cite{carpita_exploring_2019}. Therefore, knowledge about what variables lead to success on the pitch is crucial to improve the performance of the models \cite{oberstone_differentiating_2009}.

\section{Formulation} \label{sec:formulation}

The primary focus of our methodology is to predict match outcomes by forecasting the distribution of both shot quantity and quality for each game. This concept is formally captured in Equations \ref{eq:quant} and \ref{eq:qual}. Here, $\sim Quant$ and $\sim Qual$ denote the distributions of shot quantity and quality, respectively, forecasted using a machine learning model $f$. The notation $ELO_H$ and $ELO_A$ correspond to the ELO rating of the home and away teams, respectively. For simplicity, only the ELO ratings for both teams are used as inputs for this predictive model. We work under the assumption that the forecasted distributions follow the normal distribution. Therefore, we define the distribution by its mean $\mu$ and standard deviation $\sigma$. Note that, although they follow the same distribution type, the parameters that model quantity and quality are different.

\begin{equation}
\sim Quant_{H} = \mathcal{N}(\mu_{quant}, \sigma_{quant}^2), \text{ where } \mu_{quant}, \sigma_{quant} = f(ELO_{H},\ ELO_{A})
\label{eq:quant}
\end{equation}

\begin{equation}
\sim Qual_{H} = \mathcal{N}(\mu_{qual}, \sigma_{qual}^2), \text{ where } \mu_{qual}, \sigma_{qual} = f(ELO_{H},\ ELO_{A})
\label{eq:qual}
\end{equation}

With the distributions defined, we can proceed to simulate a game. For each team, either home or away, we generate a vector $hs$ or $as$. The size of this vector is defined by a sample $N$ from $\sim Quant$, with elements $S$ sampled from $\sim Qual$, as indicated in Equation \ref{eq:homeshotvector}.

\begin{equation}
\textbf{hs} = [S_1, S_2, ..., S_N],\
\text{where}\ S \sim Qual_H\ \text{and} \ N \sim Quant_H
\label{eq:homeshotvector}
\end{equation}

From these vectors, we can simulate the number of goals scored by each team in the match. For this purpose, we sample random values $X$ from a uniform distribution ranging from 0 to 1. If the random value is lower than the probability given by the $\sim Qual$, we interpret it as a goal. By adding up all the instances where a shot is considered a goal, we derive the final score for each team. Equation \ref{eq:home_goals} formulates this for the home team, and a similar process is applied to the away team.

\begin{equation}
\text{Home Goals} = \sum_{i=1}^{N} \textbf{hs}_i > X_i,\ \text{where}\ X_i \sim U(0,1)
\label{eq:home_goals}
\end{equation}

\section{Method} \label{sec:method}

\begin{figure}[h]
\centering
\includegraphics[width=0.8\textwidth]{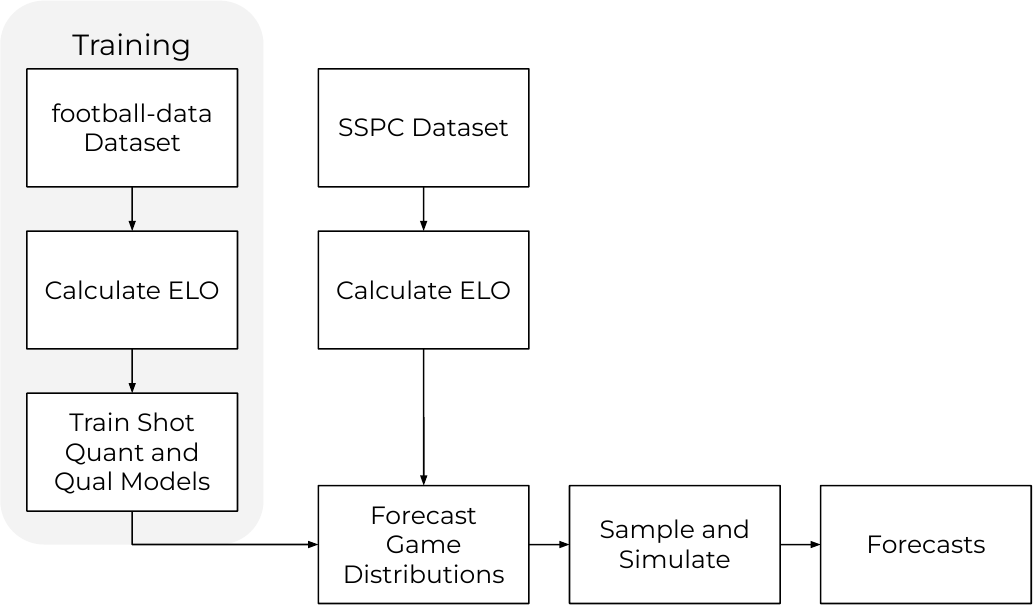}
\caption{The process we use for forecasting soccer matches. In the grey area, we include the steps to obtain the models that forecast the distribution of shot quantity and quality from the football-data.co.uk dataset. The remaining steps calculate the distributions for the games we want to forecast and then obtain the results forecasts through simulation.}\label{fig:methods}
\end{figure}

Our forecasting method comprises the processes illustrated in Figure \ref{fig:methods}, which blends machine learning techniques and ELO ratings. These steps revolve around calculation, modeling, and simulation to accurately predict soccer matches.

\subsection{Data}

The data used to train this study's models primarily came from the football-data.co.uk dataset, as the challenge dataset lacked information about shots. This dataset focuses on European leagues, providing a comprehensive overview of team performances, including the number of shots each team takes. To evaluate the performance of our models, we designated the last 300 games in the dataset for validation purposes. The main objective of this data split is to train the models that predict shot quantity and quality.

\subsection{ELO Ratings}
The ELO rating system, originally devised by Arpad Elo~\cite{arpad_e_elo_proposed_1967}, is a recognized method for assessing relative skill levels in two-player games. The first step in our methodology is to calculate these ratings for each soccer team.

The ELO rating is updated based on the outcome of the match, as described in Equations \ref{eq:elo_update} and \ref{eq:elo_probability}:
\begin{equation}
ELO_{i(t+1)} = ELO_{it} + K(O_{ijt}-P_{ijt})
\label{eq:elo_update}
\end{equation}

\begin{equation}
P_{ijt} = 1/(1+10^{-(ELO_{it}-ELO_{jt})/a})
\label{eq:elo_probability}
\end{equation}

Where:

\begin{itemize}
    \item $i$ is the team for which we are calculating the ELO, and $j$ is the opposing team.
    \item $t+1$ and $t$ denote the player's ratings after and before the match.
    \item $K$ represents the maximum change in rating (in our computations, we used $K=32$).
    \item $a$ is a constant indicating the difference in ratings corresponding to a 10-fold difference in the odds of winning (in our computations, we used $a=400$).
    \item $O$ corresponds to the actual outcome of the match (1 for a win, 0.5 for a draw, and 0 for a loss).
    \item $P_{ijt}$ is the expected probability of the team $i$ winning the match against team $j$, computed using the logistic function.
\end{itemize}

All teams start with an initial ELO rating of 500. As new matches are played, the ELO rating of each team is updated, adapting dynamically to the teams' performance over time.

\subsection{Prediction Pipeline}
The following list of steps describes the process of forecast from our approach. Note that it might differ from the mathematical explanation from Section \ref{sec:formulation} since we introduce some data manipulation techniques that allow for faster simulation when implemented. Nonetheless, the resulting probabilities are the same.
\begin{enumerate}
    \item Retrieve the ELO rating for each team. We use the last 10 games mean and standard deviation to build a distribution of ELO ratings for the team. This allows us to capture the uncertainty of the level of play of the team. For example, a lower average/higher variance team can occasionally have a higher ELO rating than a high average/low variance team.
    \item Fill a matrix of size (n\_simulations, 50) with samples from the shot quality distribution.
    \item For each row of the matrix, which represents a game simulation, sample from the shot quantity distribution. For matrix elements with an index equal to or greater than the number of shots, these elements are set to zero. This process represents the stochastic nature of the number of shots taken in a game.
    \item Within each game simulation, simulate the shots against a uniform distribution between 0 and 1. If the sample from the uniform distribution is less than the sample from the shot quality distribution, then the value is set to 1, representing a goal. Otherwise, it is set to zero, indicating no goal.
    \item Sum the values for each team to obtain the simulated score for the game. This sum reflects the final score based on the simulated shots and goals.
    \item To forecast the odds, we aggregate all game simulations. The probability of each outcome is computed as the number of times the outcome occurred divided by the number of simulations. This aggregation provides the probability distribution over possible match outcomes.
    \item For correct score predictions, we have three options: we can either use the mean and round the result, use the median (which also requires rounding for edge cases), or use the mode. These different approaches give us flexibility in determining the most probable scoreline.
\end{enumerate}

This structured pipeline allows for simulations that account for the inherent uncertainty and variability in soccer matches. By adopting a probabilistic approach, we capture a broad range of potential outcomes, which provides a more nuanced view of possible match results compared to single-value predictions. Furthermore, our approach allows the forecast of multiple types of markets. From the frequency of results, we obtain the probabilities of outcomes (task 2), correct scores (task 1), number of goals in a match, and winning by different margins, among others.

\subsection{Metrics}
The evaluation of our forecast model primarily revolves around two metrics: the Ranked Probability Score (RPS) and Root Mean Squared Error (RMSE).

The RPS is a metric that addresses a significant nuance in soccer forecasts. For example, a forecast predicting a home win that results in a draw can be considered more accurate than a forecast predicting a home win that results in an away win. Traditional metrics like accuracy do not fully capture this nuance. Therefore, RPS is a useful measure as it gives a more holistic assessment of the forecast accuracy by considering the ordinal nature of the match outcomes.

On the other hand, RMSE is a less-than-ideal metric for our purposes. RMSE measures forecast quality based on the square of the difference between the actual and forecasted values. An illustrative example can be seen in Table \ref{tab:RMSE_example}, where the square error is shown for predictions between 0, 1, 2, and 3 goals against the actual result. Our data shows that around 85\% of home team and 90\% of away team scores fall within the range of 0 to 2 goals. The RMSE metric encourages a "safe play" from the models, as forecasting 1 provides the smallest maximum error (1) within that range. On the other hand, if the forecast is 0 or 2, the error can be 4x larger than the maximum error for forecasting 1 in that range. Thus, RMSE does not necessarily accurately reflect prediction merit but benefits the “safe play” of forecasting a 1-1 result.

\begin{table}[ht]
\centering
\begin{tabular}{c|cccc}
\hline
\textbf{Forecast} & \multicolumn{4}{c}{\textbf{Actual Result}}\\
\hline
 & \textbf{0} & \textbf{1} & \textbf{2} & \textcolor{lightgray}{\textbf{3}} \\
\hline
\textbf{0} & 0 & 1 & 4 & \textcolor{lightgray}{9} \\
\textbf{1} & 1 & 0 & 1 & \textcolor{lightgray}{4} \\
\textbf{2} & 4 & 1 & 0 & \textcolor{lightgray}{1} \\
\textcolor{lightgray}{\textbf{3}} & \textcolor{lightgray}{9} & \textcolor{lightgray}{4} & \textcolor{lightgray}{1} & \textcolor{lightgray}{0} \\
\hline
\end{tabular}
\caption{Quadratic errors for various forecasts and actual results.}
\label{tab:RMSE_example}
\end{table}

Due to the limitations of RMSE, we also present our results in terms of Mean Absolute Error (MAE). Unlike RMSE, MAE provides a more intuitive measure of average forecast error, as it directly corresponds to the average absolute difference between the predicted and actual values without giving disproportionate weight to large errors. As a baseline, we provide the results of forecasting every game to end in a 1-1 draw. The baseline RMSE is 1.68, and the MAE is 1.73.

\subsubsection{Betting Rentability}

From a market perspective, comparing our estimated odds against those offered by bookmakers provides a tangible value assessment for our methods. This comparison, however, necessitates the definition of a deployment strategy. The deployment strategy's aim is to identify specific market subsets where our models can potentially outperform the market. Identifying such subsets requires considering various parameters, including the market subset definition and the margin of safety. Although this paper does not delve into these intricate details, we define two strategies to provide an evaluation of the overall profitability of our model:

\begin{enumerate}
\item \textbf{Strategy 1:} Bet one unit on each game if the odds offered by the bookmaker are higher than our estimate.
\item \textbf{Strategy 2:} Bet an amount equivalent to the inverse value of the forecasted odds if the odds offered by the bookmaker are higher than our estimate.
\end{enumerate}

It is important to note that returns are not guaranteed; the model requires consistent monitoring and updating as market conditions are continuously evolving. The baseline for comparison is the bookmaker's margin, which is around -5.6\% in the test set. The RPS for bookmakers is approximately 0.198. These figures serve as the benchmarks for assessing the profitability of our model.

\section{Experiments} \label{sec:experiments}
\subsection{Modeling Shot Quantity and Quality}

The core of our approach involves modeling the distributions that underpin our methods. To this end, we constructed models utilizing the Home Team ELO and Away Team ELO as input variables. The outputs are twofold: (1) the total number of shots in a match or (2) the estimated goals per shot, derived by dividing the total goals by the shot count for each game.

We chose a simple input feature set with just the ELO ratings because they succinctly encapsulate the comparative strength of each team. However, we understand that some uncorrelated factors like match importance, rest days between matches, and the league’s playing intensity can be used to improve our approach.

Modeling the quantity and quality of shots as a distribution (mean and standard deviation) instead of a single value is crucial. It allows us to represent the inherent uncertainty in soccer matches, reflecting the fact that the same teams playing under the same conditions could have different outcomes. For example, assume that a team can be represented by a normal distribution of goals scored $(\mu, \sigma)$. A team with parameters (1, 0) will never win against a team with parameters (2, 0). But if the team has parameters (1, 0.5) against a team (2, 0) will win \~0.13\% of the encounters. Moreover, if the team (1, 0.5) faces a team (2, 0.5), it will win \~2.70\% of the time. Variance has a significant impact on the forecasts of soccer matches. 

We investigated two approaches to predict the distribution: (1) a neighbor-based approach and (2) a supervised learning approach. Both approaches assume that shot quality and quantity follow a normal distribution.

The neighbor-based approach calculates the neighbors for the game we want to forecast and computes the prediction as the average of the neighbors' values. It computes uncertainty as the standard deviation of the neighbors' values. Euclidean distance is used to determine the neighbors. 

\begin{figure}[h]
\centering
Linear Regression
\includegraphics[width=\textwidth]{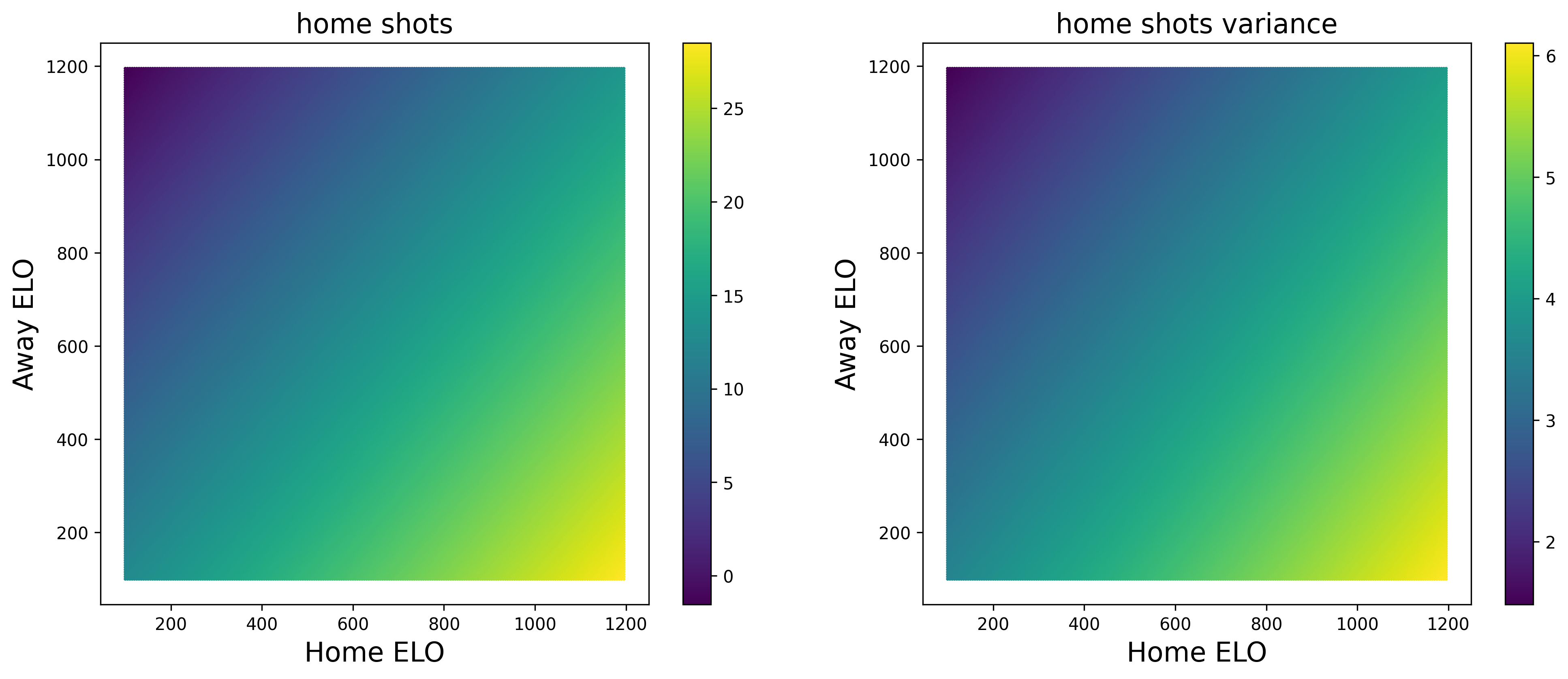}
K-Nearest Neighbors
\includegraphics[width=\textwidth]{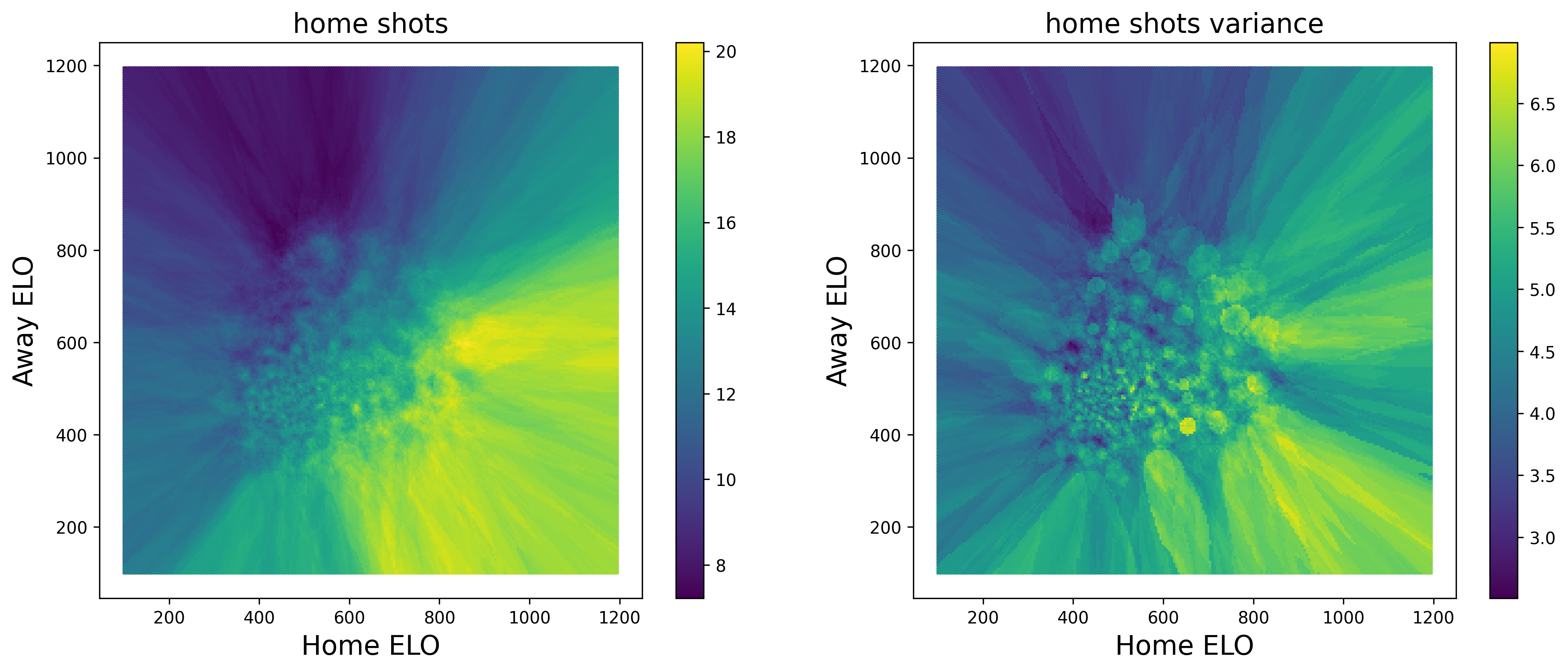}
\caption{The prediction surface of the shot quantity models using Linear Regression and K-Nearest Neighbors.}\label{fig:shots_gradient}
\end{figure}

The second approach is a supervised learning model that predicts the distribution. The process is twofold: first, we train a model on the label to compute the mean. Subsequently, we compute uncertainty by creating another model that uses the error from the first model's forecasts to the actual values in the training set to predict the standard deviation. The forecasts of both approaches are shown in Figure \ref{fig:shots_gradient}. Note that after the prediction, we incorporate constraints related to soccer, such as requiring an integer and a positive number of shots.

Table \ref{tab:validation_results} presents the results of the experiments. We tested multiple algorithms in the supervised learning approach, like Linear Regression, Decision Tree, and Random Forest. We selected these models due to their good performance without requiring extensive hyperparameter tuning.

\begin{table}[ht]
\centering
\begin{tabular}{lccccc}
\hline
\textbf{Model} & \textbf{RPS} & \textbf{RMSE} & \textbf{RMSE} & \textbf{MAE} & \textbf{MAE}\\
\textbf{Variant} & \textbf{} & \textbf{Median} & \textbf{Mode} & \textbf{Median} & \textbf{}\\
\hline
Linear Regression & 0.215 & 1.76 & 1.78 & 1.87 & 1.82 \\
K-Nearest Neighbors & 0.213 & 1.73 & 1.85 & 1.79 & 1.87 \\
Decision Tree & 0.215 & 1.77 & 2.09 & 1.83 & 2.10 \\
Random Forest & 0.213 &	1.74 & 1.78 & 1.80 & 1.81 \\
Linear Regression* & \textbf{0.216} & - & \textbf{1.70} & - & -\\

\hline
\end{tabular}
\caption{Summary of results on our validation set.\\ *Competition results.}
\label{tab:validation_results}
\end{table}

\section{Results} \label{sec:results}

The outcomes from our analyses are summarised in Table \ref{tab:results_summary}, highlighting the results obtained from the football-data.co.uk test set. This dataset allows for an in-depth exploration of profitability metrics. 

\begin{figure}[h]
\centering
\includegraphics[width=0.9\textwidth]{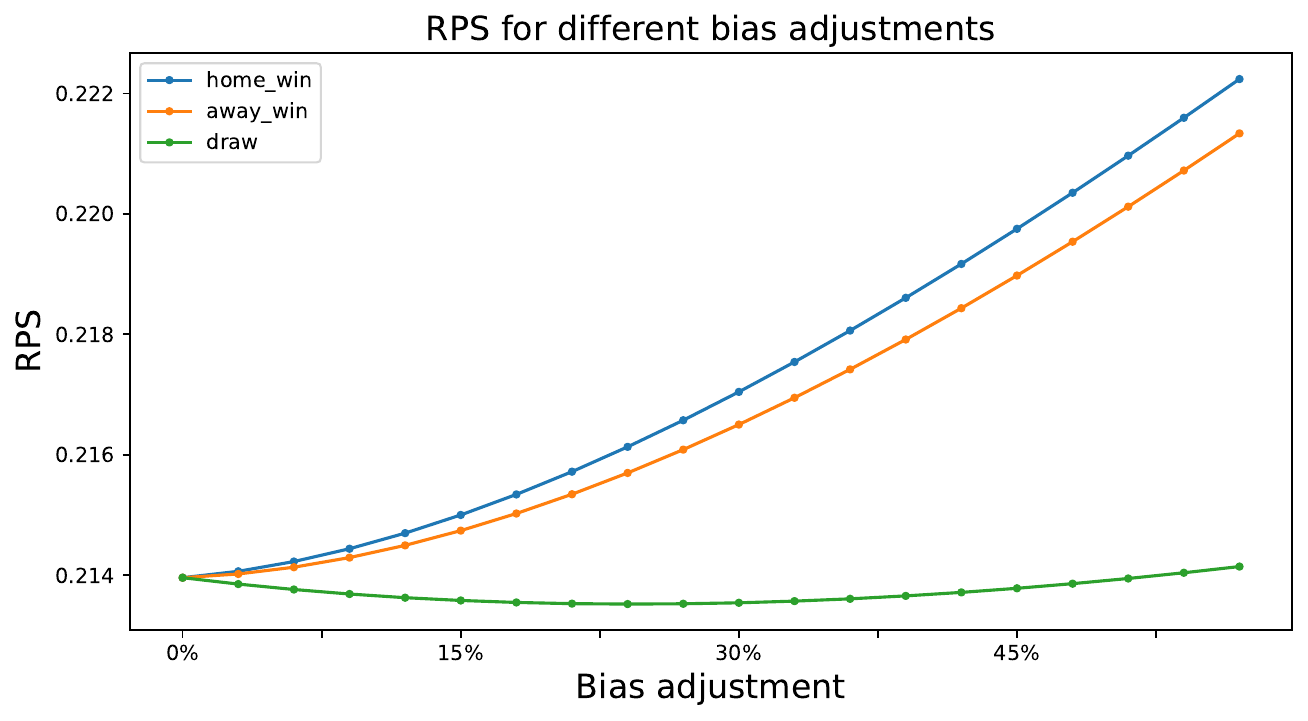}
\caption{The impact of increasing the likelihood of an outcome in the overall 
 results. Through this analysis, we observe that our approach contains a systematic bias error when forecasting the probability of a draw.}\label{fig:bias_adjustment}
\end{figure}

Furthermore, we sought to evaluate potential biases within our models. This is achieved by identifying the optimal adjustment value for each odd, as depicted in Figure \ref{fig:bias_adjustment}. This bias analysis aids in better understanding the reliability and accuracy of our forecasting models.

\begin{table}[ht]
\centering
\begin{tabular}{lcc}
\hline
\textbf{Metric} &  \multicolumn{2}{c}{\textbf{Result}} \\
\hline
RPS &  \multicolumn{2}{c}{0.201} \\
\hline
\textbf{Metric} & \textbf{Using Median} & \textbf{Using Mode} \\
\hline
RMSE & 1.58 & 1.64 \\
MAE & 1.65 & 1.66 \\
\hline
\textbf{Strategy} & \textbf{Rentability} & \textbf{Over Baseline} \\
\hline
Strategy 1 & -0.8\% & +4.8\% \\
Strategy 2 & 1.1\% &  +6.7\% \\
\hline
\end{tabular}
\caption{Summary of results on the football-data.co.uk test set. Values were obtained as an average of 50 runs on the test set.}
\label{tab:results_summary}
\end{table}

\section{Discussion} \label{sec:discussion}

Our methodology required two crucial decisions: selecting an appropriate machine learning model and choosing between utilizing median or mode for predicting correct scores.

Regarding the latter, despite Table \ref{tab:results_summary} showing a slight preference towards using the median based on the RMSE metric, the difference becomes negligible when considering the more appropriate MAE metric. This pattern persists in Table \ref{tab:validation_results}, with the RMSE favoring the median and the MAE favoring the mode. Given our perspective that the MAE better measures the merit of our model, we decided to use the mode for calculating correct scores.

The second decision involved the choice of a machine learning technique to model the relationship between ELO ratings and shot quantity and quality. We evaluated four methods: Linear Regression, Decision Trees, Random Forest, and KNN. The Random Forest model performed the best, although we employed Linear Regression for the competition submission due to timing constraints.

On the profitability aspect, our model notably outperformed the baseline, achieving a +6.7\% return with Strategy 2, enough to secure profitability. This result underlines the importance of matching the right betting strategy with the model, as Strategy 2 surpassed Strategy 1 by almost 2\%.

Bias analysis revealed an improvement in our model's performance upon increasing the likelihood of a draw by 27\%. Although this improvement is marginal, it indicates a potential avenue for further research, given the inherent difficulty in predicting draws in soccer.

Despite the constraints of the competition's scope, which limits available data to final scores, and our own deliberate simplifications, such as minimal feature engineering by using only team ELO ratings as inputs, our methodology performed commendably, remaining within a 5\% margin of the competition's top entries. The introduction of domain knowledge, not directly into the models, but in formulating the problem, resulted in a promising method for forecasting soccer matches.

\section{Conclusion} \label{sec:conclusion}

In this paper, we introduced a method that leverages game simulation for predicting soccer matches, specifically, forecasting shot quantity and quality for each team. This methodology was explored within the confines of the Springer Soccer Prediction Challenge.

Our method has proven competitive, showcasing the efficacy of combining domain-specific knowledge with the power of machine learning techniques. Working within the limitations of simplified features, we found that our approach held its ground against more intricate models. Furthermore, our model extends its utility beyond mere match result prediction; it demonstrated the capability to provide profitable insights in the betting realm, where it exhibited performance superior to the market baseline.

To conclude, this research accentuates the potential embedded in a simulation-based approach for soccer match prediction. Competitions such as the Springer Soccer Prediction Challenge provide valuable platforms to stimulate research into critical yet overlooked foundational concepts. The constraints imposed by the task compel creative exploration of the fundamental issues, thereby deepening our understanding and shaping the future of soccer prediction research.

\subsection{Future Work}

This study opens up numerous avenues for further exploration and enhancement of our soccer match forecasting model. Here are some promising directions we intend to pursue in future iterations of this work:

\begin{itemize}
  \item \textit{Feature Expansion:} We plan to incorporate additional features into our model to increase its predictive power. Notable among these are variables indicating match importance and rest days for the teams. These factors could significantly affect team performance and, consequently, the outcomes of matches.
  \item \textit{Rating Systems:} Different rating systems might yield different insights and predictive accuracy. Therefore, experimenting with other rating systems apart from ELO could be beneficial.
  \item \textit{Betting Strategies:} A more in-depth study to find the betting strategies where our model outperforms the market is very likely to enhance the model's profitability. Identifying the specific market subsets and conditions that favor our model could lead to more refined betting strategies.
  \item \textit{Modeling Techniques:} Given the complexity of soccer matches, linear models may not fully capture the intricacies of shot quantity and quality. We intend to explore more non-linear machine learning methods to potentially improve these aspects of the model, especially as we incorporate more features.
  \item \textit{Data Availability:} Access to expected goals data could allow us to construct a more sophisticated and accurate shot quality model. The potential to incorporate such data in future iterations is an exciting prospect.
\end{itemize}

By addressing these areas, we hope to improve our model's performance and profitability, further enhancing its applicability in sports forecasting.

\backmatter

\bmhead{Acknowledgments}

This work is financed by National Funds through the Portuguese funding agency, FCT - Fundação para a Ciência e a Tecnologia, within project UIDB/50014/2020.

\section*{Declarations}

\bmhead{Funding}

This work is financed by FCT - Fundação para a Ciência e a Tecnologia.

\bmhead{Conflict of interest}
The authors have no conflict of interest to declare that are relevant to the content of this article.

\bmhead{Code availability}
The code and the models are available at 
https://github.com/nvsclub/SpringerSoccerChallengeCode.

\bmhead{Data availability}
The data is publicly available to reproduce this work, and should be obtained from the original sources:
\begin{itemize}
    \item https://football-data.co.uk/
    \item https://sites.google.com/view/2023soccerpredictionchallenge
\end{itemize}

\bmhead{Ethics approval}
Not applicable. The research conducted does not require ethics approval.

\bmhead{Consent to participate}
Not applicable. This manuscript does not conduct any research with individuals.

\bmhead{Consent for publication}
Not applicable. The authors authored all the content in the manuscript.

\bmhead{Authors' contributions}
All authors contributed to the ideas presented in this paper through proposals or refinement of existing ideas; the same can be said about the methodology used in this project. The code implementations are the responsibility of Tiago Mendes-Neves and Yassine Baghoussi. Tiago Mendes-Neves provided the writing of the first draft of this article, to which all authors contributed with revisions and recommendations. Luís Meireles, Carlos Soares, and João Mendes-Moreira were responsible for supervising the research activity and mentorship.

\bibliography{sn-bibliography}

\end{document}